\documentclass[12pt]{spieman}  
\newcommand{\LP}[1]{${\rm LP_{#1}}$}

\usepackage{amsmath,amsfonts,amssymb}
\usepackage{graphicx}
\usepackage{setspace}
\usepackage{tocloft}
\usepackage{tabularray}
\usepackage{lineno}

\title{On the Potential of Spectroastrometry with Photonic Lanterns}

\author[a]{Yoo Jung Kim}
\author[a]{Michael P. Fitzgerald}
\author[a]{Jonathan Lin}
\author[b]{Yinzi Xin}
\author[c]{Daniel Levinstein}
\author[c]{Steph Sallum}
\author[b]{Nemanja Jovanovic}
\author[d]{Sergio Leon-Saval}

\affil[a]{Department of Physics \& Astronomy, University of California, 430 Portola Plaza, Los Angeles, CA 90095, USA}
\affil[b]{Department of Astronomy, California Institute of Technology, 1200 E. California Blvd., Pasadena, CA 91125, USA}
\affil[c]{Department of Physics \& Astronomy, University of California Irvine, 4129 Frederick Reines Hall, Irvine, CA 92697, USA}
\affil[d]{Sydney Astrophotonic Instrumentation Laboratory, School of Physics, The University of Sydney, Sydney, NSW 2006, Australia}

\cftpagenumbersoff{figure}
\cftpagenumbersoff{table} 
\graphicspath{{./}{figures/}}

\begin{document} 
\maketitle

\providecommand\mnras{Monthly Notices of the Royal Astronomical Society}
\providecommand\apj{The Astrophysical Journal}
\providecommand\apjl{The Astrophysical Journal, Letters}
\providecommand\apss{Astrophysics and Space Science}
\providecommand\aap{Astronomy \& Astrophysics}
\providecommand\pasp{Publications of the Astronomical Society of the Pacific}

\begin{abstract}
We investigate the potential of photonic lantern (PL) fiber fed spectrometers for two-dimensional spectroastrometry. Spectroastrometry, a technique for studying small angular scales by measuring centroid shifts as a function of wavelength, is typically conducted using long-slit spectrographs. However, slit-based spectroastrometry 
requires observations with multiple position angles to measure two-dimensional spectroastrometric signals.
In a typical configuration of PL-fed spectrometers, light from the focal plane is coupled into the few-moded PL, which is then split into several single-mode outputs, with the relative intensities containing astrometric information. The single-moded beams can be fed into a high-resolution spectrometer to measure wavelength-dependent centroid shifts. We perform numerical simulations of a standard 6-port PL and demonstrate its capability of measuring spectroastrometric signals. The effects of photon noise, wavefront errors, and chromaticity are investigated. When the PL is designed to have large linear responses to tip-tilts at the wavelengths of interest, the centroid shifts can be efficiently measured. Furthermore, we provide mock observations of detecting accreting protoplanets. PL spectroastrometry is potentially a simple and efficient technique for detecting spectroastrometric signals.
\end{abstract}


\section{Introduction}\label{sec:intro}

High angular resolution enables the detailed study of a variety of objects such as young stellar objects, jets, circumstellar environments, and quasars. Spectroastrometry is a method to study angular scales smaller than the point-spread function (PSF) size for objects whose morphology changes with wavelength (such as those exhibiting emission lines), by measuring the relative position of an unresolved object as a function of wavelength\cite{bec82, bai98, whe08}. 
A wavelength-dependent shift in the center of light --- the definition of spectroastrometric signal --- may indicate the presence of a companion, outflow, or any spatially extended feature, revealing details about the structure within the PSF (seeing- or diffraction-limit). 
The spectroastrometry method has been used for many purposes, such as studying young binaries \cite{bai98b}, circumstellar disks \cite{pon08, bri15}, kinematics of stellar outflows \cite{whe04}, and the broad-line region of quasars \cite{ste15, bos21}. With diffraction-limited PSFs enabled by adaptive optics (AO), sub-milliarcsecond precisions can be achieved with 10\,m-class telescopes in the near-infrared.

Spectroastrometry leverages high spectral resolution to resolve kinematic structures at small angular scales or to detect companions with narrow spectral lines.
Therefore, long-slit echelle spectrometers are typically used for spectroastrometric observations, as they can achieve high spectral resolution and spatial sampling along one axis. However with long-slit spectrographs, at least two observations with different slit orientations are required in order to obtain two-dimensional spectroastrometric signals. Another downside of using a slit is that a distorted PSF or uneven illumination can introduce an artificial spectroastrometric signal \cite{bra06, whe15}. At medium resolutions, integral field spectrometers can enable two-dimensional spectroastrometry with decreased artifacts, although with an increased complexity \cite{dav10, got12, mur13}. 

\begin{figure}
    \centering
    \includegraphics[scale=0.45]{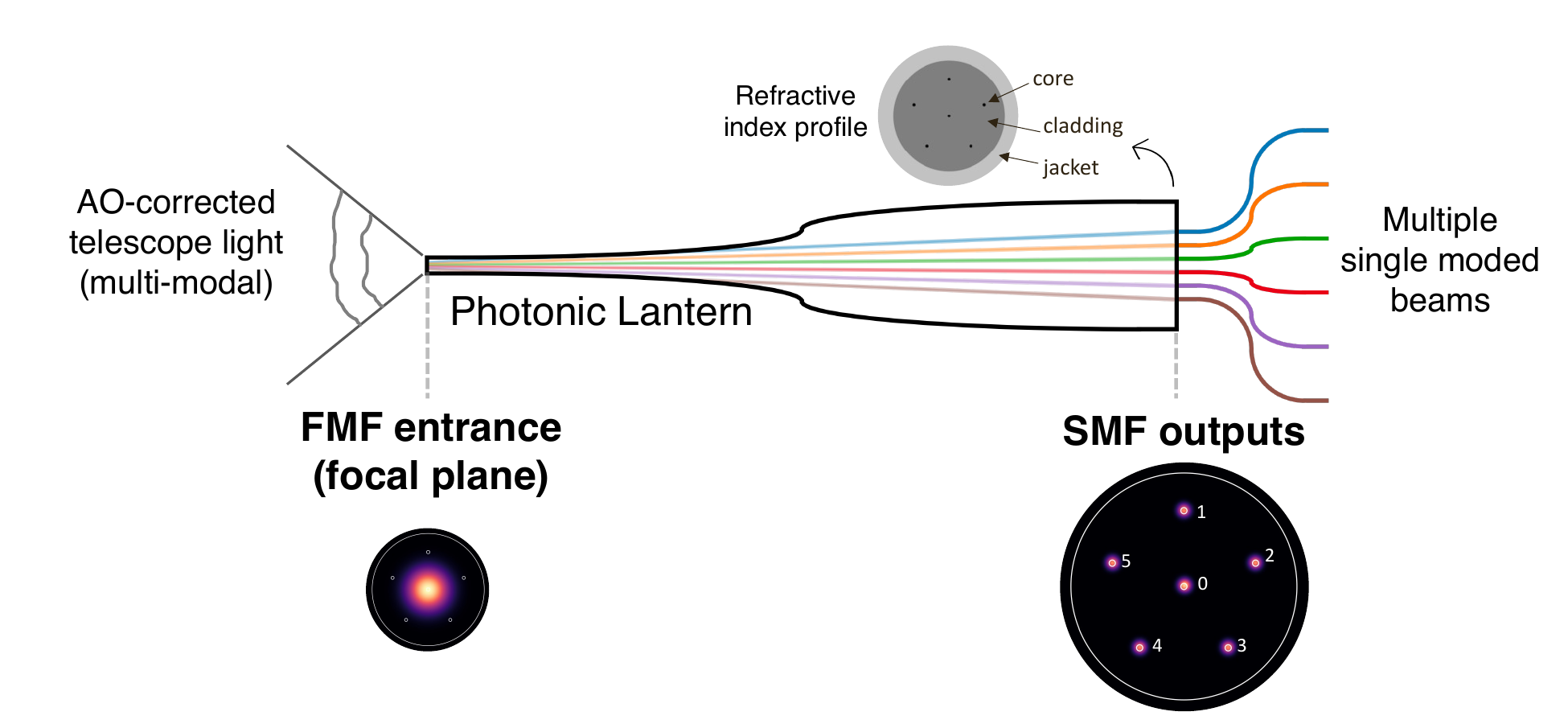}
    \caption{A schematic diagram of the standard 6 port photonic lantern (PL) used for our simulation. The inset figure positioned above the PL shows the refractive index profile of the PL at the few-core fiber (FCF) end. The telescope light is coupled in the FMF end, propagates through the lantern, and becomes effectively confined in the SMFs. The light coupled in the SMFs can be fed to a spectrometer for spectroscopy. The relative spectra in output SMFs can be used to measure the center of light as a function of wavelength (i.e., spectroastrometry.)}
    \label{fig:PL}
\end{figure}

In this study, we explore the potential of using photonic lanterns (PLs) for two-dimensional spectroastrometry \cite{kim22, levinstein23}. PLs are tapered waveguides that gradually transition from a few-mode fiber (FMF) or multi-mode fiber (MMF) geometry to a bundle of single-mode fibers (SMFs) \cite{leo10, leo13, bir15} (Figure \ref{fig:PL}). When the AO-corrected telescope light couples into the FMF end of the PL in the focal plane, it becomes confined within the SMF cores as it propagates through the lantern. Thus, a PL converts multimodal telescope light into multiple single-moded beams. The light in the SMFs can be fed into spectrometers for spectroscopy. PLs are potentially advantageous for diffraction-limited precision high-resolution spectroscopy due to the spatial filtering nature of the SMFs \cite{jov16} and the high coupling efficiency compared to coupling light directly into SMFs \cite{lin21} (AO-assisted SMF-fed spectrometers such as KPIC \cite{delorme21}, IRD \cite{kotani18}, PARVI \cite{gibson20}, iLocator \cite{crass16}, HiRISE \cite{vigan24}, and HISPEC/MODHIS \cite{mawet19}). Potential uses of PLs for astronomical observations have been studied in context of nulling \cite{xin22} and coherent imaging \cite{kim24}.

The dispersed SMF outputs from a single PL placed at the focal plane can act like a small integral field unit (IFU), sensing spatial features in very small angular scales. Recent studies on PLs as focal plane wavefront sensors (PLWFS) \cite{cor18, nor20, lin22, lin23, lin23b} have demonstrated that PL output intensities can be used to sense low-order aberrations. The capability of PLWFS is related to PL spectroastrometry, since centroid shifts correspond to tip-tilt Zernike modes, the lowest modes excluding the unsensed piston mode. With PL output spectra, wavelength-dependent two-dimensional centroid shifts can be efficiently measured, eliminating the need for slicing or resampling of the focal plane fields as in IFU and using only a few single-moded spectral traces. Thus, both high angular resolution and high spectral resolution can be achieved with PLs.

This paper is structured as follows. In \S\ref{sec:concept}, we show how spectroastrometric signals can be recovered from PL outputs and present numerically simulated spectroastrometric signals for a 6-port PL. In \S\ref{sec:errors}, we discuss sources of errors in spectroastrometric signal recovery: photon noise, static wavefront errors (WFEs), and time-varying WFEs.  
In \S\ref{sec:chromaticity}, we address the chromatic behaviors of PLs relevant to spectroastrometry. In \S\ref{sec:mockobs}, we present mock observations of accreting protoplanets. In \S\ref{sec:discussion} we discuss benefits of PL spectroastrometry and considerations on PL design.

\section{Concept}\label{sec:concept}

\subsection{Spectroastrometric signals for PLs}\label{ssec:sasignal}

Consider an instantaneous monochromatic wavefront incident on the telescope pupil $\textbf{E}_{\rm p}$ (represented as an $M$-dimensional complex-valued vector of field samples) and its Fourier transform, the focal plane electric field $\textbf{E}_{\rm f}$. An $N$-port PL couples $\textbf{E}_{\rm f}$ at the PL entrance. Following the analytical model of Lin et al. 2022 \cite{lin22}, the instantaneous PL output field can be described as
\begin{equation}
    \textbf{E}_{\rm SMF} = A\, \textbf{E}_{\rm p}.
\end{equation}
The $A$ matrix is the complex-valued transfer matrix which includes propagation from the pupil to the focal plane and the lantern transition, with the dimension of $N \times M$ for $M$ pupil samples. $\textbf{E}_{\rm SMF}$ is the $N$-dimensional vector, the complex amplitudes in output SMFs. Thus, the $A$ matrix maps the pupil plane wavefront to PL outputs.

The transfer matrix $A$ corresponds to the inverse of the $M \times N$ matrix formed by stacking $N$ number of $M$-dimensional pupil plane PL principal modes (PLPMs \cite{kim24}). Numerically, the $i$-th mode pupil plane PLPM is computed by backpropagating the fundamental mode of the $i$-th output SMF to the pupil plane. The $i$-th row of the $A$ matrix corresponds to $i$-th effective pupil function of the $i$-th output SMF. 
Then the intensities in the output SMFs can be expressed as
\begin{equation}\label{eq:intensity}
    \mathbf{I}_{\rm SMF} = |A\, \textbf{E}_{\rm p}|^2.
\end{equation}
We abbreviate $\mathbf{I}_{\rm SMF}$ for $\mathbf{I}$ afterwards for simplicity and write $i$-th SMF output intensity as $\mathbf{I}_i$.

An input scene, consisting of incoherent sources (excluding coherent sources such as masers), 
can be described as an incoherent sum of $n$ point sources located at angular coordinates $\boldsymbol{\alpha}_l = (x_l, y_l)^{\rm T}$ and weighted by the flux factor $f_l$, with $l=0, 1, ..., n-1$. The wavefront from $l$-th point source is written as a tilted plane wave, $\textbf{E}_{{\rm p}}^l = {f_l}^{1/2} \exp{(i R_{\alpha} \boldsymbol{\alpha}_l)}$, where $R_{\alpha}$ is the $M \times 2$ basis matrix whose columns are pupil plane tip-tilt basis vectors. 
The output intensities are then modeled as
\begin{equation}\label{eq:intensity_total}
    \mathbf{I}_{\rm tot} = \sum_{l=0}^{n-1} \mathbf{I}^l = \sum_{l=0}^{n-1} |A\, \textbf{E}_{{\rm p}}^l|^2.
\end{equation}

\begin{figure}[hbt!]
    \centering
    \includegraphics[width=1\linewidth]{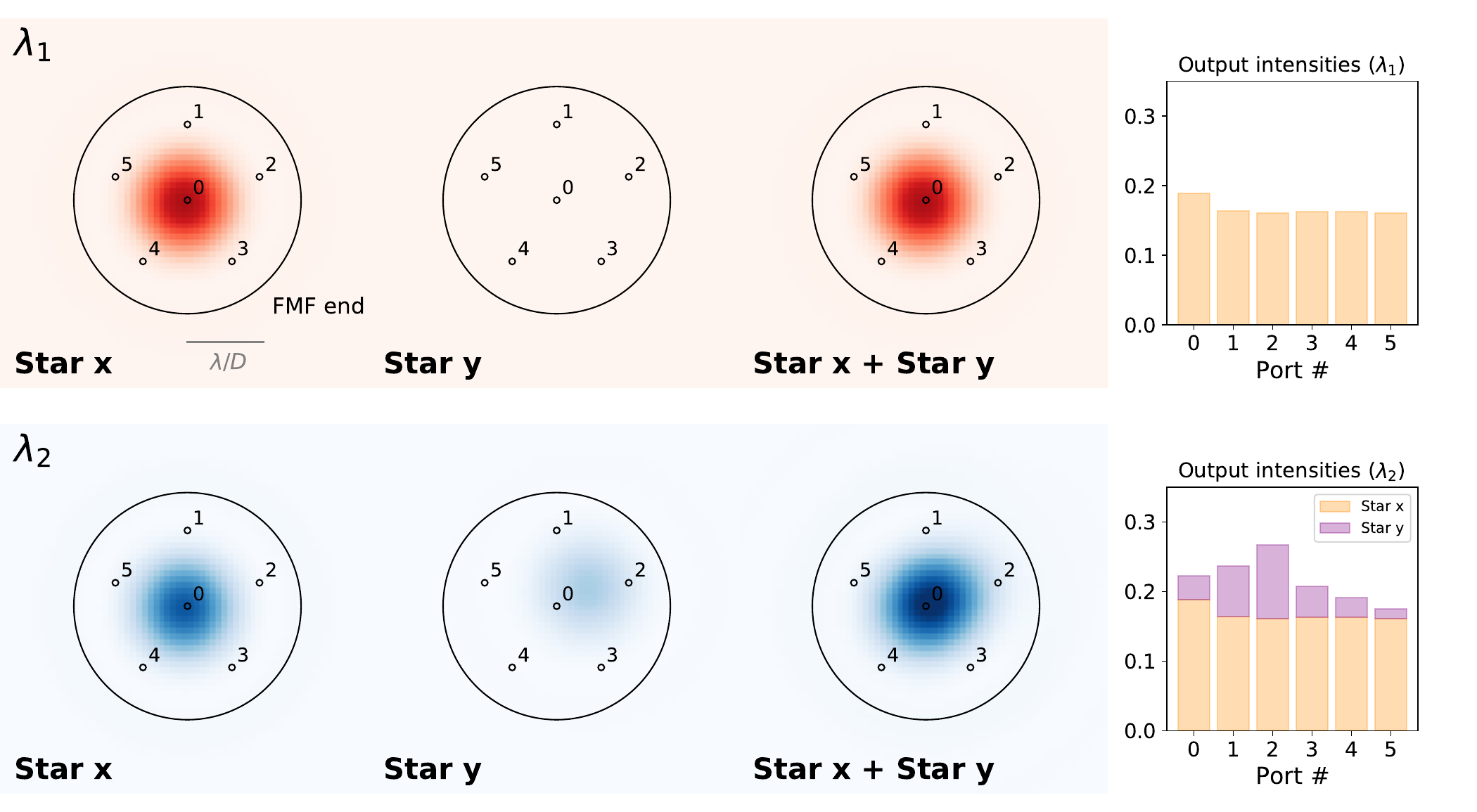}
    \caption{(Left) Focal plane intensity maps for a simple example case, a bright star without spectral features (star $x$) located at the center and an off-center fainter star (star $y$) with an emission line at $\lambda_2$. Top rows are for $\lambda_1$ and bottom rows are for $\lambda_2$. The FMF end geometry of a standard 6-port PL is displayed as black circles. Due to the emission line of star $y$ at $\lambda_2$, the center of light of the sum of the two stars is shifted in $\lambda_2$ while in $\lambda_1$ does not show any shift. The center of light shift is smaller than the FMF radius (order of $\lambda/D$) and should be sensed by the lantern as a tip-tilt aberration. (Right) Simulated output intensities for $\lambda_1$ (top) and $\lambda_2$ (bottom). At $\lambda_2$, the flux of the emission line from the star $y$ is added onto the flux of the star $x$, altering the relative intensities.}
    \label{fig:demo_SA_focal}
\end{figure}

Figure \ref{fig:demo_SA_focal} illustrates a simple example case of binary point sources, star $x$ and star $y$. Star $x$ dominates the total flux at wavelength $\lambda_1$, but star $y$ has an emission line at wavelength $\lambda_2$. Due to the emission line, the center of light shifts towards the star $y$ at wavelength $\lambda_2$ which is the spectroastrometric signal. Our goal is to detect the star $y$ using spectroastrometry on this emission line. On the right we display the simulated $\mathbf{I}_{\rm tot}$ for a standard 6-port PL. Details on the simulation are described in \S\ref{sssec:sim}. Although the on-axis light from the star $x$ (orange) couples to ports 1 through 5 evenly, the off-axis light from the star $y$ (purple) couples to the port 2 the most. The relative intensity profiles are determined by the separation, contrast, and position angle (PA) of the star $y$. Thus, the difference in output relative intensities on and off the emission line gives the two-dimensional spectroastrometric signal.

Let us consider an input scene with a small angular extent ($\ll\lambda/D$). 
Assuming small $\boldsymbol{\alpha}_l$, the wavefront can be approximated as $\textbf{E}_{{\rm p}}^l \approx {f_l}^{1/2} (\textbf{1} + i R_{\alpha} \boldsymbol{\alpha}_l)$. Following Lin et al. 2022 \cite{lin22}, the intensities in the output SMFs from each point source can be expressed as
\begin{equation}\label{eq:linear_expansion}
    \mathbf{I}^{l} = |A\, \textbf{E}_{{\rm p}}^l|^2 \approx f_l (|A \textbf{1}|^2 + B R_{\alpha} \boldsymbol{\alpha}_l),
\end{equation}
where $B$ is the linear response matrix defined as
\begin{equation}
    B_{ij} \equiv 2 {\rm Im} \Bigl[ A_{ij}^* \sum_k A_{ik} \Bigr].
\end{equation}
We define $B' \equiv B R_\alpha$, the $B$ matrix projected onto tip-tilt basis, which describes the linear intensity responses to tip-tilt. 
Then the total intensity is the sum of the $n$ point sources:
\begin{align}\label{eq:intensum}
    \begin{aligned}
        \mathbf{I}_{{\rm tot}} = \sum_{l=0}^{n-1} \mathbf{I}^l 
        &\approx \left(\sum_{l=0}^{n-1} f_l\right) \left( |A \textbf{1}|^2 + B' \frac{\sum\limits_{l=0}^{n-1} f_l \boldsymbol{\alpha}_l}{\sum\limits_{l=0}^{n-1} f_l} \right) \\
        &\approx f_{\rm tot} \bigl( |A \textbf{1}|^2 + B'  \boldsymbol{\alpha}_{\rm centroid} \bigr).
    \end{aligned}
\end{align}
$f_{\rm tot} \equiv \sum\limits_{l=0}^{n-1} f_l$ and $\boldsymbol{\alpha}_{\rm centroid}$ is defined as the center of light angular coordinate vector. 
This suggests that if the angular size of the on-axis object is small that the tip-tilts corresponding to the angular size perturb lantern output intensities linearly, the intensity responses mainly describe the center of light shift of the input scene.  

In practice, $f_{\rm tot}$ in each wavelength bin is an unknown factor. 
Even if the exact spectra of the object is known a priori, there may still be chromaticity in how the light couples into the PL due to stochastic processes, such as atmospheric dispersion.
Instead, we can define {\it normalized intensities} in the output SMFs ($\textbf{I}_{n}$) as the output intensities divided by the total sum of the output intensities, $\sum\limits_{i=0}^{N-1} \textbf{I}_{i}$. The normalized intensities can be linearized similarly in this regime such as
\begin{equation}
    \textbf{I}_{{n}}^l \approx \textbf{I}_{n0} + B'_n \boldsymbol{\alpha}_l
\end{equation}
where $\textbf{I}_{\rm n0}$ is the normalized intensity for the on-axis point source, $|A \textbf{1}|^2 / \sum\limits_{i=0}^{N-1} |A \textbf{1}|_i^2$ and $B'_n$ is the normalized intensity linear response matrix, $B' / \sum\limits_{i=0}^{N-1} |A \textbf{1}|_i^2$.
Then the equation \ref{eq:intensum} can be rewritten as
\begin{equation}
    \textbf{I}_{{n, {\rm tot}}} \approx \textbf{I}_{n0} + B'_n \boldsymbol{\alpha}_{\rm centroid}
\end{equation}
in the linear intensity response regime. 

Finally, inverting the above equation, the center of light --- the definition of spectroastrometric signals --- can be recovered from the observed $\Delta\textbf{I}_n \equiv \textbf{I}_{{n, {\rm tot}}} - \textbf{I}_{n0}$ as:
\begin{equation}\label{eq:centroid}
    \boldsymbol{\alpha}_{\rm centroid} \approx B'^{+}_n \Delta\textbf{I}_n.
\end{equation}
$B'^{+}_n$ is the left pseudo-inverse of the $B'_n$ matrix. Note that both $B'_n$ and $\textbf{I}_{\rm n0}$ are determined by the lantern design, the transfer matrix $A$. They can also be empirically determined in a lab with calibration light sources.

Note that the $R_{\alpha}$ matrix can be extended to an $M \times m$ matrix $R$ to include additional modes ($m$) beyond tip and tilt. Then $B' = B R$ describes general linear intensity responses to  aberrations. An $N$-port PL can sense up to $m=N-1$ non-piston Zernike modes \cite{lin22}. Using the generalized linear intensity response matrix and calculating the right-hand side of the Equation \ref{eq:centroid}, one obtains an $m$-dimensional vector, with first two elements corresponding to tip-tilt. If the intensity responses were perfectly linear, the entries other than tip-tilts should be zero. Any deviations from the linear approximation would manifest as nonzero values in the other modes (modal confusion). We provide examples in \S\ref{sssec:nonlinear}. Also, the effects of static and time-varying WFEs over an exposure can alter the intensity response matrix, which we discuss in \S\ref{ssec:staticWFE} and \S\ref{ssec:varyingWFE}.

\subsection{Simulated spectroastrometric signals}\label{ssec:simulatedsignal}

In this subsection, we conduct numerical simulations of a standard 6-port lantern ($N=6$) and simulate spectroastrometric signals for several simple astronomical scenes. 

\subsubsection{Simulation}\label{sssec:sim}

\begin{figure}[hbt!]
    \centering
    \includegraphics[width=1\linewidth]{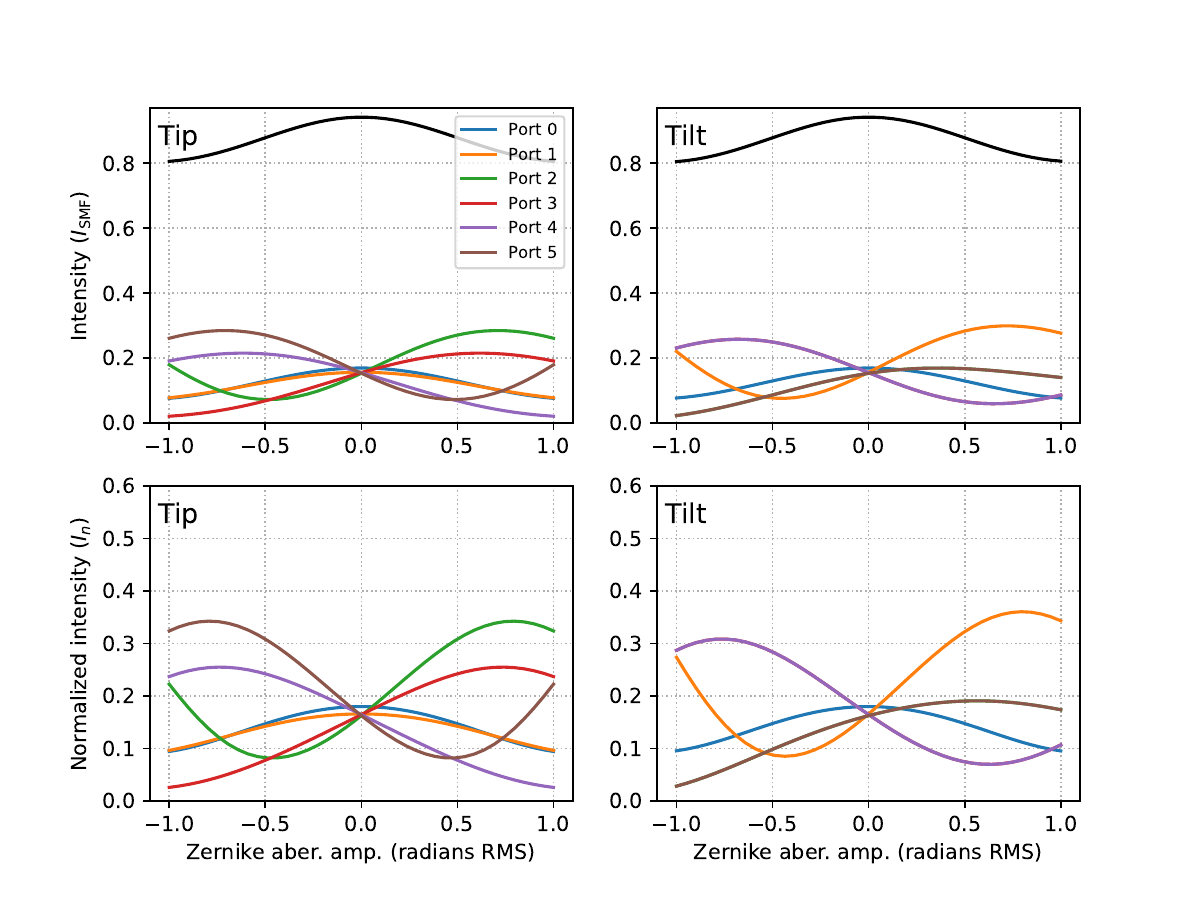}
    \caption{(Top) Simulated SMF output intensities ($\mathbf{I}_{\rm SMF}$) for a standard 6-port lantern, as a function of tip-tilt Zernike mode aberration amplitudes. The total intensity incident on the pupil is set to unity. The black lines denote the total intensity in the 6 output SMFs. (Bottom) Same as the top panels but the intensities are normalized ($\mathbf{I}_{\rm n}$) by the sum of the total intensities (the black lines). See Figure \ref{fig:PL} for the geometry of the PL used for the simulation and \S\ref{sssec:sim} for the simulation setup.}
    \label{fig:linearity}
\end{figure}

The simulated lantern is a standard 6-port PL ($N=6$) of which all the SMFs have the same core radius and refractive index, with 5 cores arranged symmetrically around a central core (Figure \ref{fig:PL}). We assume the cladding index of 1.444, cladding-jacket index contrast of $5.5\times 10^{-3}$, and core-cladding index contrast of $8.8\times10^{-3}$. Each SMF core diameter is chosen to be 4.4 \textmu m and the lantern entrance diameter to 20 \textmu m. The lantern taper length is set to 2 cm and the taper scale is set to 8. We numerically backpropagate fundamental modes in the SMFs to the lantern entrance using the \texttt{lightbeam}\cite{lin21lightbeam} python package to determine focal-plane PLPMs.

For the telescope, we assume an unobstructed circular aperture $D$ of 10~m. The simulations are monochromatic with wavelength 1.55~\textmu m ($\lambda/D\sim 32$~mas).
The focal plane PLPMs are backpropagated to the telescope pupil to determine the pupil plane PLPMs and the transfer matrix $A$ using the \texttt{HCIPy} (High Contrast Imaging for Python) package \cite{por18}. The focal length is optimized to maximize the total coupling (intensity coupled over all of the supported LP modes) of an unaberrated on-axis point source.

The top rows of Figure \ref{fig:linearity} display the simulated output intensities of a single point source ($\mathbf{I}$) as a function of tip (left) and tilt (right). In the small tip-tilt regime, the intensity responses are linear for most of the ports. The black curves indicate the total intensity over the 6 output SMFs, the total coupling efficiency. The bottom row shows the normalized output intensities ($\mathbf{I}_{n}$), which also behave linearly for small tip-tilts, but more non-linearly in larger tip-tilt regimes. The $y$-intercepts correspond to $\mathbf{I}_{\rm n0}$ and the slopes at the origin are the $B'_n$.

\subsubsection{Example scenes}\label{sssec:scenes}

Using the simulated lantern transfer matrix, we simulate spectroastrometric signals for simple scenes. The lantern output spectra are calculated as an incoherent sum of output spectra from point sources that constitute the input scene (Equation \ref{eq:intensity_total}). We neglect the chromaticity of the lantern's transfer matrix for this simulation and assume that the transfer matrix is constant over the wavelength range, [1.51, 1.59] \textmu m. We discuss the effects of chromaticity in \S\ref{sec:chromaticity}. 

\begin{figure}[hbt!]
    \centering
\includegraphics[width=1\linewidth]{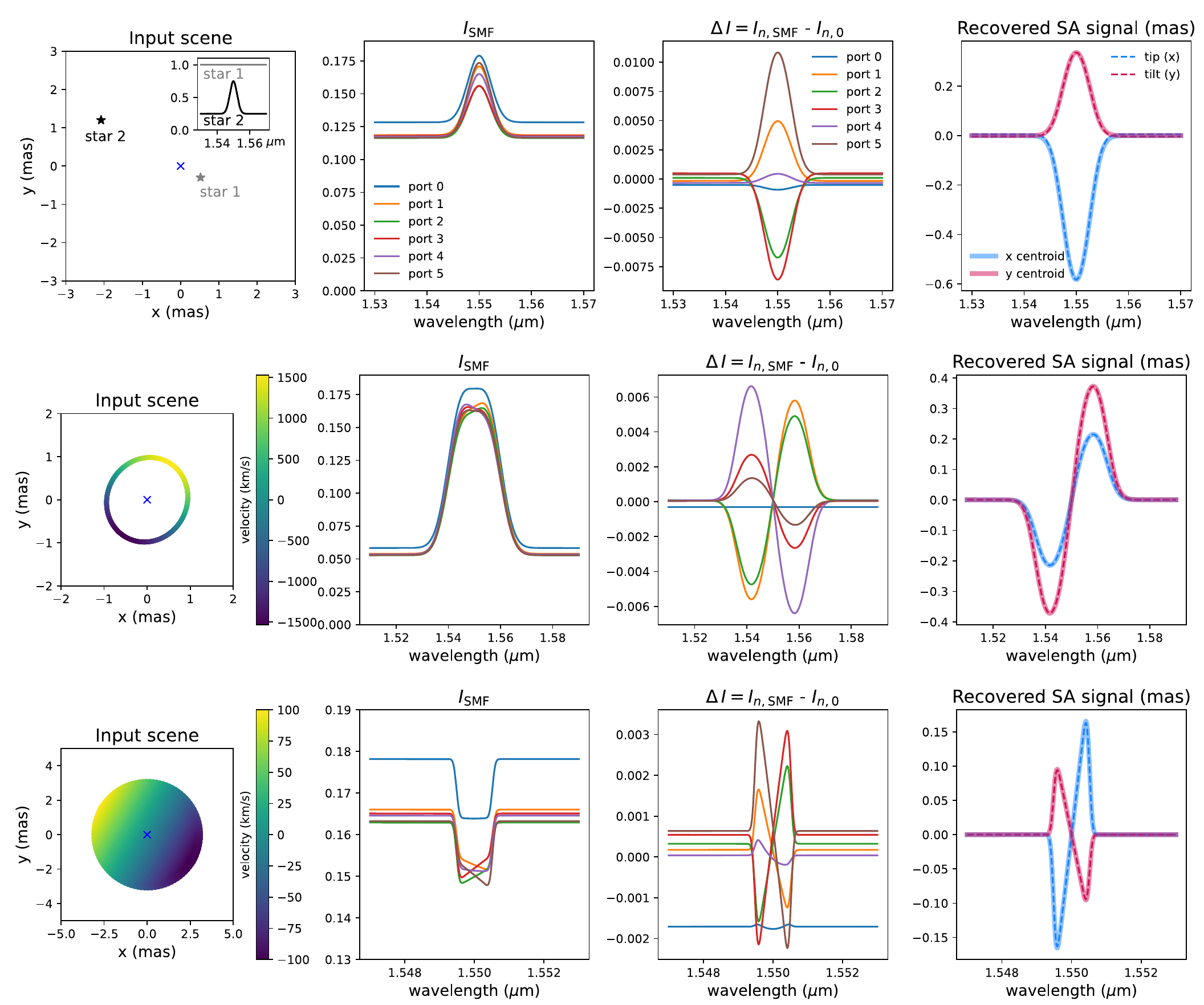}
\caption{Example PL spectroastrometric signals for three simple scenes, (top) binary point sources with an emission line on one, (middle) a rotationally-broadened emission line originating from a ring structure such as broad-line region of a quasar, and (bottom) a rotationally-broadened absorption line from a resolved star. The first column illustrates the input scenes. The second column and the third column display simulated output spectra and normalized spectra, respectively, of a standard 6-port PL. Due to the positional variation of the line-emitting region, the PL output spectra show variations in fluxes across the ports. This variation can be used to recover the center of light as a function of wavelength, the fourth column.}
\label{fig:scenes}
\end{figure}

Top panels of Figure \ref{fig:scenes} show a case for binary point sources separated by 3 mas. The left panel displays the input scene, one star with a constant continuum emission (gray) and the other star of 4:1 line-continuum flux ratio, exhibiting an emission line of FWHM = 1,000 km s$^{-1}$ at 1.55~\textmu m (black). The center of continuum emission is located at the origin (blue cross). The second panel from the left shows the lantern output spectra of each SMF, $\textbf{I}_{\rm tot}$. The output spectra are normalized by the peak of the summed spectra (summed over the ports). The amplitude of the emission line varies from port to port. The third panel presents $\Delta\textbf{I} = \textbf{I}_{{n,{\rm tot}}} - \textbf{I}_{\rm n0}$, the normalized spectra (in each wavelength bin) subtracted by the on-axis point source normalized intensity. The fourth panel shows the recovered spectroastrometric signals $\boldsymbol{\alpha}_{\rm centroid}$ as dashed lines, Equation \ref{eq:centroid}. The true center of light positions as a function of wavelength are overlaid as thick solid lines. This shows that the two-dimensional centroid shift on the emission line can be recovered with the simple linear reconstruction method.

Middle panels of Figure \ref{fig:scenes} show a case for a rotating ring ($v_{\rm rot} = 4,000$ km s$^{-1}$) of 1 mas radius, such as broad-line region of quasars \cite{ste15, bos21}. Each flux element shows a broad emission line of FWHM = 2,000 km s$^{-1}$ (local broadening). The inclination angle is 22.5 degrees and the PA is 60 degrees. We assume continuum level of 25\% for the amplitude of the emission line. Since the blueshifted and redshifted flux of the emission line originate from different locations of the ring, sinusoidal spectroastrometric signals are expected. If the line is sufficiently resolved in the spectral axis, PL spectroastrometry can provide estimation on the two-dimensional centroid shifts which can then be used to infer the kinematic structures, potentially useful for constraining more complicated models.

Bottom panels of Figure \ref{fig:scenes} display a case for a rotating sphere of 3 mas radius with an absorption line, such as a rapidly rotating star ($v_{\rm rot} = 100$ km s$^{-1}$). The star is modeled as a collection of 2,500 light-emitting patches on a hemisphere, with an absorption line of FWHM = 30 km s$^{-1}$. Similarly to the ring case, the two-dimensional centroid shift can be detected and the stellar rotation such as spin axis can be characterized with the spectroastrometric signals.

\subsubsection{Nonlinear intensity response regime}\label{sssec:nonlinear} 

\begin{figure}[hbt!]
    \centering
\includegraphics[width=1\linewidth]{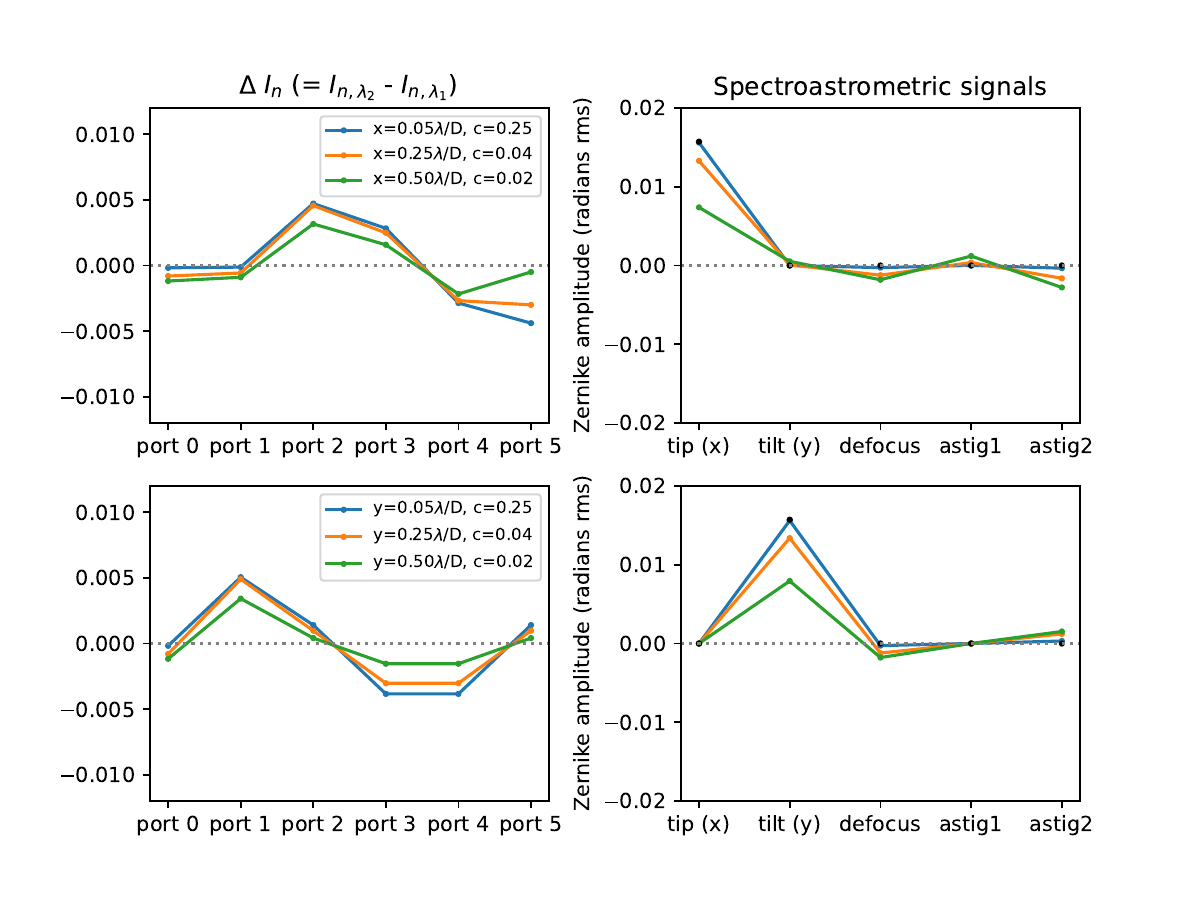}
\caption{Differences in normalized intensities ($\Delta \mathbf{I}_{\rm n}$; left) and recovered spectroastrometric signals ($\boldsymbol{\alpha}$; right) for the binary model. The companion with emission line is placed along the $x$-axis (top) and $y$-axis (bottom), respectively, with three different separation and contrast values. On the right panel, the true expected spectroastrometric signals are indicated as black dots.}
\label{fig:binaryexample}
\end{figure}

If the object is more extended such that the linear intensity response approximation (Equation \ref{eq:linear_expansion}) is no longer valid, the estimation of the light centroid by linear approximation (Equation \ref{eq:centroid}) fails. This effect is illustrated in Figure \ref{fig:binaryexample}, which presents simulated spectroastrometric signals for binary models with the same centroid shifts but with different contrasts and separations. Since a standard 6-port PL is sensitive to the first five (non-piston) Zernike modes, tip-tit, defocus, and astigmatism \cite{lin22}, the $R_{\alpha}$ matrix in Equation \ref{eq:linear_expansion} can be extended to an $M \times 5$ matrix whose columns are the five mode basis vectors. Then the right hand side of Equation \ref{eq:centroid} becomes a five-element vector, the Zernike aberration amplitudes in tip-tilt, defocus, and astigmatism. If the input scene has a small angular extent then Equation \ref{eq:linear_expansion} is valid, the first two entries should represent the centroid position in tip-tilts and the other three entries should equal to zero. However, if the output intensities deviate from the linear approximation, the right hand side of Equation \ref{eq:centroid} will show modal confusion, detecting non-zero signals in defocus and astigmatism modes. For binary models, as shown in the right panels of Figure \ref{fig:binaryexample}, increasing the binary separation results in non-zero signals in defocus and astigmatism modes and underestimation of centroid shifts (tip-tilts). 

\begin{figure}
    \centering
\includegraphics[width=1\linewidth]{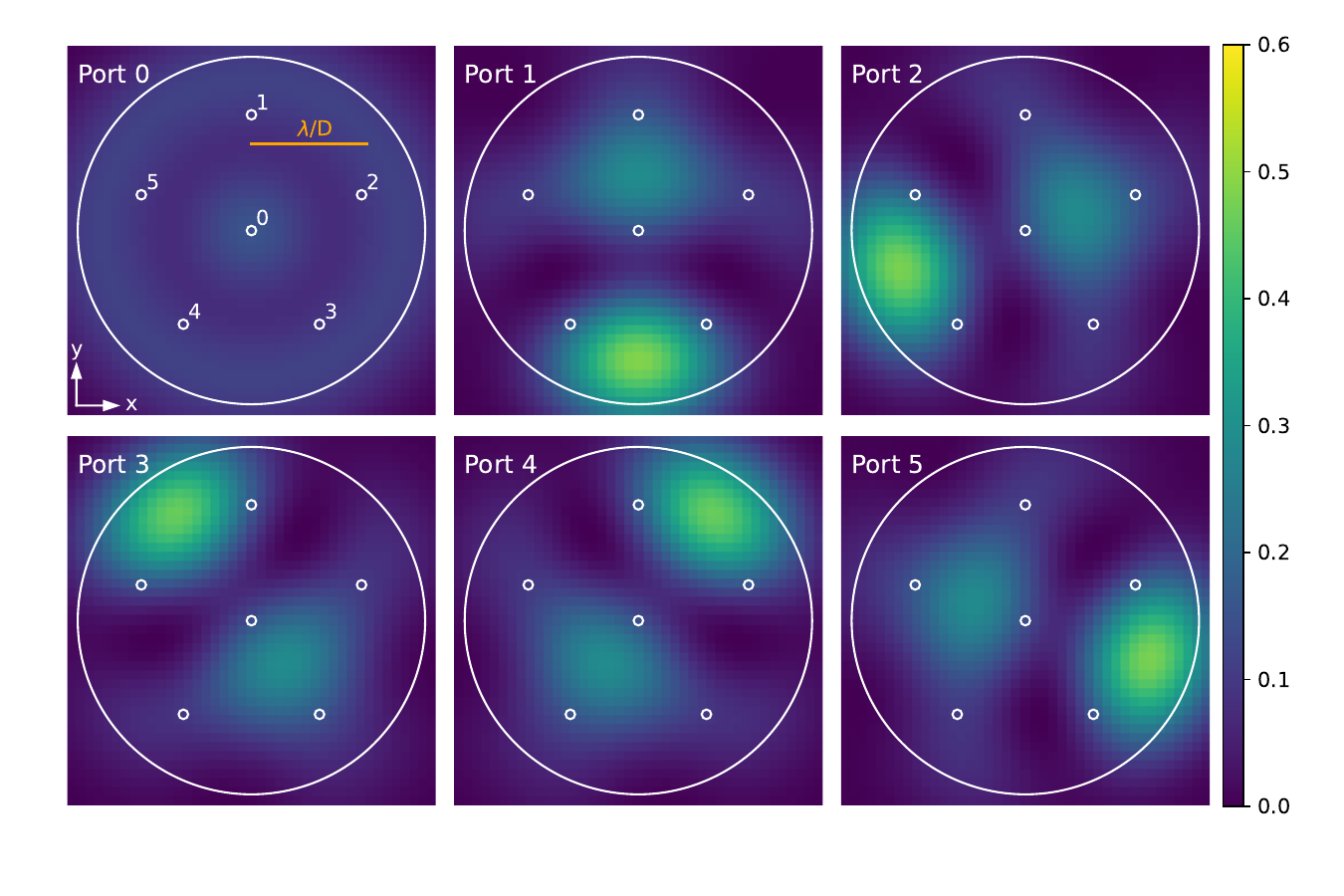}
\caption{Simulated intensity responses in each SMF port of the 6-port PL, as a function of the x, y position of a point source at the focal plane. The FMF end geometry is plotted as white circles, of which the diameter is 20 \textmu m. The total intensity of the point source is set to 1. Intensities in each port are sensitive to the position of the point source, so the relative intensities between the ports can be used to determine the separation and the PA of the companion.}
\label{fig:coupling}
\end{figure}

In practice, when non-negligible spectroastrometric defocus and astigmatism signals are detected, it indicates that the angular scale of the intensity distribution is larger than the linear intensity response regime (unless there is chromatic WFE). 
It is the regime where more information other than the centroid shifts can be extracted. In this case, models of output intensities given an input scene can be generated with knowledge of the transfer matrix $A$ or using an empirical PL coupling map -- output intensities in each port scanned over a grid of $x$, $y$ PL positions in the focal plane \cite{lin22spie}. Figure \ref{fig:coupling} shows the simulated coupling maps: intensity responses as a function of $x$, $y$ position of a point source at the focal plane. The gradients evaluated at the center of the map correspond to the $B'$ matrix, the linear intensity response. An input scene that best describes the measured $\Delta \mathbf{I}_n$ can be constrained using the transfer matrix or the coupling maps.
An example of this approach will be discussed in \S\ref{sec:mockobs} as case B for a binary system.

\section{Errors in PL spectroastrometric signal recovery }\label{sec:errors}

In \S\ref{sec:concept}, we showed simulated spectroastrometric signals for an ideal case. 
In this section, we investigate major error sources that one needs to consider in spectroastrometric observations with PLs: photon noise and WFE. Photon noise and instantaneous WFE due to turbulence are stochastic processes that contribute to random errors. Systematic WFEs and time-averaged WFE effects averaged over an exposure introduce 
systematic errors that require calibration.

\subsection{Photon noise}\label{ssec:photnoise}

For PLs, the amplitude of the intensity response to tip-tilt (the $B'_n$ matrix) determines how efficient PLs are for detecting spectroastrometric signals. In the photon noise-limited regime, conventional spectroastrometric accuracy (uncertainties in centroid measurement) scales as 
\begin{equation}\label{eq:sig_conventional}
    \sigma_{\rm centroid} \sim \frac{\sigma_{\rm sig}}{\sqrt{N_{\rm phot}}} = \frac{\rm FWHM}{2.355 \sqrt{N_{\rm phot}}}
\end{equation}
where FWHM is the full-width at half maximum ($\sim \lambda/D$ if diffraction-limited) and $N_{\rm phot}$ is the total number of detected photons \cite{whe08}. For the PL spectroastrometry, the errors in $i$-th Zernike mode phase recovery in the linear regime scale as
\begin{equation}\label{eq:sig_PL}
    \sigma_{i}^2 \approx \frac{{ \sum\limits_{j=1}^{N} (B'^{+}_{n,ij})^2 \textbf{I}_{\rm n0,j}
    }}{{N_{\rm phot}}}.
\end{equation}
The numerator is determined by the transfer matrix $A$, which is an analog of the Gaussian width ($\sigma_{\rm sig}$, the standard deviation of the Gaussian PSF profile) in the Equation \ref{eq:sig_conventional} for tip-tilt modes. This effective resolution is a function of the inverse of the linear intensity response matrix $B'_n$ and can be interpreted as a metric describing spectroastrometric sensitivity. 
This equation is verified through numerical simulations with Poisson noise.
For our simulated lantern, the value of the numerator is 
0.82 Zernike rms amplitudes in radians, or 0.52$\lambda/D$. This is comparable to the Gaussian width 0.43$\lambda/D$ at the diffraction limit, implying that photon noise-limited errors of PL spectroastrometry are comparable to those of conventional spectroastrometry.

\subsection{Static WFEs}\label{ssec:staticWFE}
\begin{figure}[hbt!]
    \centering
    \includegraphics[width=1\linewidth]{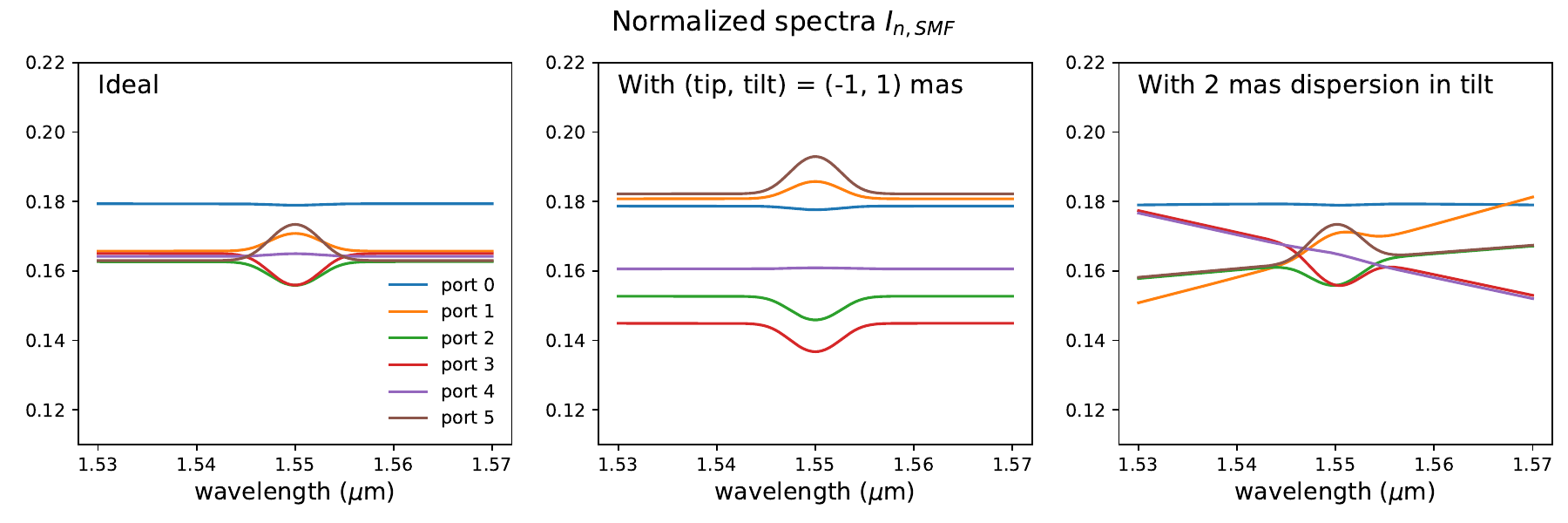}
    \caption{Example normalized spectra for binary point sources described in the top row of Figure \ref{fig:scenes}, (left) ideal case without systematics, (middle) with systematic tip-tilt, and (right) with centroid shift in tilt varying linearly with wavelength. The continuum levels can be used to eliminate the systematic effects.}
    \label{fig:normspec_systematic}
\end{figure}

Static WFEs can arise from a variety of reasons, such as optical aberrations, misalignments, atmospheric dispersion, and AO-residuals. They alter the intensity responses for an on-axis point source ($\textbf{I}_{\rm n0}$) and linear intensity responses $B'_n$.  

Figure \ref{fig:normspec_systematic} shows the normalized spectra $\textbf{I}_{\rm n}$ for the binary example case shown in the top panels of Figure \ref{fig:scenes}. In the left the ideal case without any systematics is displayed. The continuum levels of port 1 through 5 are nearly equal, but the emission line at $\lambda=1.55$ \textmu m exhibits spectroastrometric signals. Middle panel shows the case with systematic tip-tilt displacement of (-1, 1) mas. The misalignment affects both the continuum levels and the emission line. Note that tip-tilt misalignment shifts the origin in the coupling map (Figure \ref{fig:coupling}) for $\textbf{I}_{\rm n0}$ and $B'_n$. The continuum normalized intensities correspond to the shifted $\textbf{I}_{\rm n0}$. 
Besides tip-tilt misalignment, higher-order static WFEs that are not filtered by the PL (defocus and astigmatism modes for a 6-port PL) can introduce similar effects, affecting the continuum and the emission line equally. Unsensed higher-order modes have much smaller effects. 

Right panel shows the case with wavelength-dependent tilt displacement, for example caused by residuals in differential atmospheric dispersion correction. We assumed linear dispersion of 2 mas in the wavelength range [1.53, 1.57] \textmu m. If the spectral line of interest has smaller width than the scale of wavelength-dependent systematics, the slow variations in continuum levels can be used to eliminate those effects and recover spectroastrometric signals.

\subsection{Time-varying WFEs averaged over an exposure}\label{ssec:varyingWFE}

\begin{figure}[hbt!]
    \centering
    \includegraphics[width=1\linewidth]{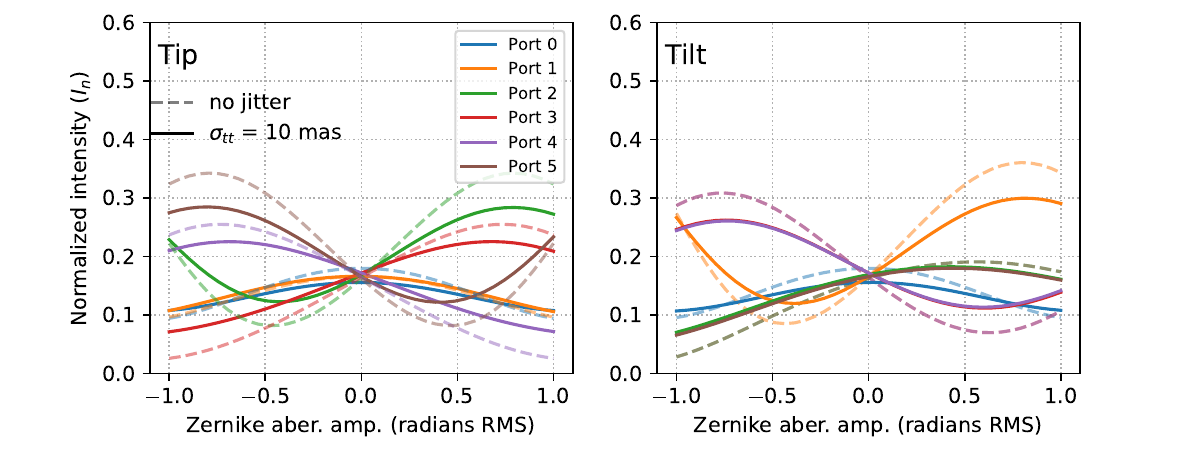}
    \caption{
    Same as bottom panels of Figure \ref{fig:linearity}, but with  10\,mas of tip-tilt jitter. The normalized intensity responses without jitter are represented as dashed lines. The intensity responses to tip-tilt are systematically decreased with jitter.}
    \label{fig:linearity_jitter}
\end{figure}

A more complicated effect is in the time-varying WFEs averaged over a long exposure, which can systematically affect spectroastrometric signal recovery. Consider a simple case with tip-tilt jitter only.
Let us assume that random samples of the tip-tilt errors follow a multivariate normal distribution
\begin{equation}
    \begin{pmatrix}
        \delta x \\ \delta y
    \end{pmatrix}
    \sim
    \mathcal{N}
    \left( 
    \begin{pmatrix}
        0 \\ 0
    \end{pmatrix},
    \begin{pmatrix}
        \sigma_x^2 & \sigma_{xy} \\
        \sigma_{xy} & \sigma_y^2
    \end{pmatrix}
    \right).
\end{equation}
Due to nonlinearities in the intensity responses, the $B'_n$ matrices are modified. Figure \ref{fig:linearity_jitter} shows the intensity responses to tip-tilt modified by the 10 mas of tip-tilt jitter ($\sigma_{x} = \sigma_{y} = 10$\,mas, $\sigma_{xy} = 0$). For reference, the intensity responses without jitter are shown as dashed lines. The linear slopes $B'_n$ are systematically modified by jitter as follows:
\begin{equation}\label{eq:jitter_cubic}
    \begin{aligned}
        \tilde{B}'_{n,ix} &\approx B'_{n,ix} + (3D'_{n,ixxx} \sigma_{x}^2+ 2D'_{n,ixxy} \sigma_{xy} + 
        D'_{n,ixyy} \sigma_{y}^2) \\
        \tilde{B}'_{n,iy} &\approx B'_{n,iy} + (3D'_{n,iyyy} \sigma_{y}^2 + 2D'_{n,ixyy} \sigma_{xy} + D'_{n,ixxy} \sigma_{x}^2) 
    \end{aligned}
\end{equation}
where $i$ is the port index and $D'_n$ is the normalized cubic intensity response tensor (see Lin et al. 2022\cite{lin22} for cubic expansion of PL intensity responses). This shows that the more cubic PL intensity responses are, the more susceptible the lantern linear responses are to the tip-tilt jitter. With time-varying WFEs in sensed higher-order modes (defocus and astigmatism for a 6-port PL), more cubic terms are added in Equation \ref{eq:jitter_cubic}.

In practice, the bias in the spectroastrometric signals due to the change in $B'_n$ may be calibrated using various approaches, such as 
by taking empirical $B'_n$ measurement on the target on-sky or a calibrator observation. Moreover, the fact that the transfer matrices vary slowly on wavelength (Section \ref{sec:chromaticity}) may be used to determine the modified $B'_n$ matrix, assuming that the optical path difference is achromatic. We defer developing practical methods in presence of realistic WFEs to future work.

\section{Chromatic behavior of PL principal modes}\label{sec:chromaticity}

In the previous sections, we simulated spectroastrometric signals using the transfer matrix computed for $\lambda = 1.55$ \textmu m for certain specified PL design parameters. However, PLPMs are in fact expected to vary slowly with wavelength. In this section, we discuss the PL chromaticity and its implications on spectroastrometry.

Let us consider light propagation from one of the SMFs (fundamental mode) to the multimode lantern end. Along the transition region, the single fundamental mode propagating in the SMF will eventually spread out of the core and couple to the neighboring SMF cores leading to a set of supermodes (the eigenmodes of the system, representing the collective behavior of the coupled cores).\cite{leo10} This set of supermodes will become non-degenerate with distinct propagation constants at the multimode end of the lantern transition. Hence, the single input fundamental mode evolves into these non-degenerate supermodes and reaches the other end (the FMF entrance) as an unique orthogonal solution leading to the PLPM. Thus, the PLPM is a superposition of orthogonal modes, which are approximately the $N$ LP modes (for step-index PLs with circular entrances). Due to the propagation constant difference ($\Delta \beta_{i,j}$) between the non-degenerate supermodes $i$ and $j$, the relative phase between the supermodes evolves as they propagate. The relative phase between the supermodes $i$ and $j$ is imprinted in the PLPMs as
\begin{equation}\label{eq:relphase}
    \Delta \varphi_{i,j} = \int_0^{z_{\rm taper}} \Delta \beta_{i,j}(z) dz
\end{equation}
where the ${z_{\rm taper}}$ is the taper length. Since the propagation constant $\beta$ relies on wavelength, $\Delta \varphi_{i,j}$ also changes with wavelength, $\Delta \varphi_{i,j} = \Delta \varphi_{i,j} (\lambda)$. This causes chromatic behavior of PLPM patterns, and consequently on the on-axis normalized intensities ($\mathbf{I}_{\rm n0}$) and intensity response matrices ($B'_n$).

\begin{figure}
    \centering
\includegraphics[width=1.0\linewidth]
{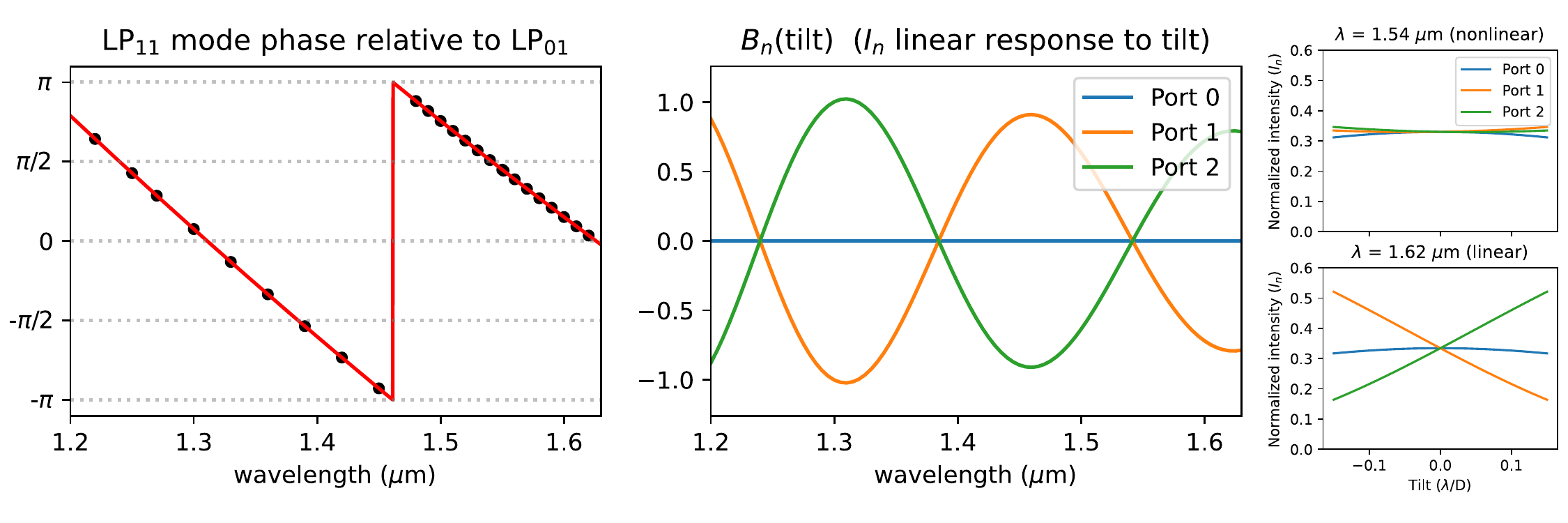}
\caption{(Left) Relative phase between \LP{01} and \LP{11} modes found from chromatic simulations of a standard 3-port PL. Black dots indicate simulated points and red curve shows the second-order polynomial fit. (Middle) Tilt components of the normalized intensity response matrix (rad$^{-1}$), showing oscillating patterns as a function of wavelength. This is related to the relative phase between the two LP modes. (Right) Normalized intensities as a function of tilt, for two wavelengths. At 1.54 \textmu m the LP modes are 90 degrees out of phase, the corresponding PLPM is symmetric, and intensity responses to tilt aberration are nonlinear. At 1.62 \textmu m the LP modes are in phase, the PLPM is asymmetric, and intensity responses are linear.} 
\label{fig:chromaticity}
\end{figure}

Figure \ref{fig:chromaticity} shows an example for a 3-port PL. We performed numerical back propagation of single-moded beams of several different wavelengths from the output SMFs to the lantern entrance and found the \LP{01} and \LP{11} mode phases of the PLPMs. The simulated 3-port PL is similar to the 6-port PL but has a smaller entrance core size (6 \textmu m) and three residual (non-guiding) cores forming an equilateral triangle instead of six. As can be seen from the left panel, the relative phase between \LP{01} and \LP{11} modes varies slowly as a function of wavelength, with a period of about 0.3 \textmu m. In the middle panel we display the tilt component of the $B'_n$ matrix, the linear intensity response given tilt aberration. The port 0 is not sensitive to tilt due to the geometry of our simulated PL. The responses of port 1 and port 2 oscillate as a function of wavelength. For the 3-port PL, we find that intensity responses behave linearly at wavelengths where \LP{01} and \LP{11} modes are in phase (multiples of $\pi$). If they are $\pi/2$ out of phase, the PLPMs are symmetric, lacking sensitivity to asymmetries, and showing zero linear response to tip-tilts ($B$ matrix). Examples for the two wavelengths are shown on the right panels. Note that the period of chromatic behavior depends on the taper length, which is fixed for a fabricated device. However, it offers another degree of freedom during the lantern fabrication process to tailor the chromatic behavior. More phase difference is accumulated if the lantern is longer, making the tip-tilt response oscillation period shorter. 

\begin{figure}[hbt!]
    \centering
\includegraphics[width=0.8\linewidth]
{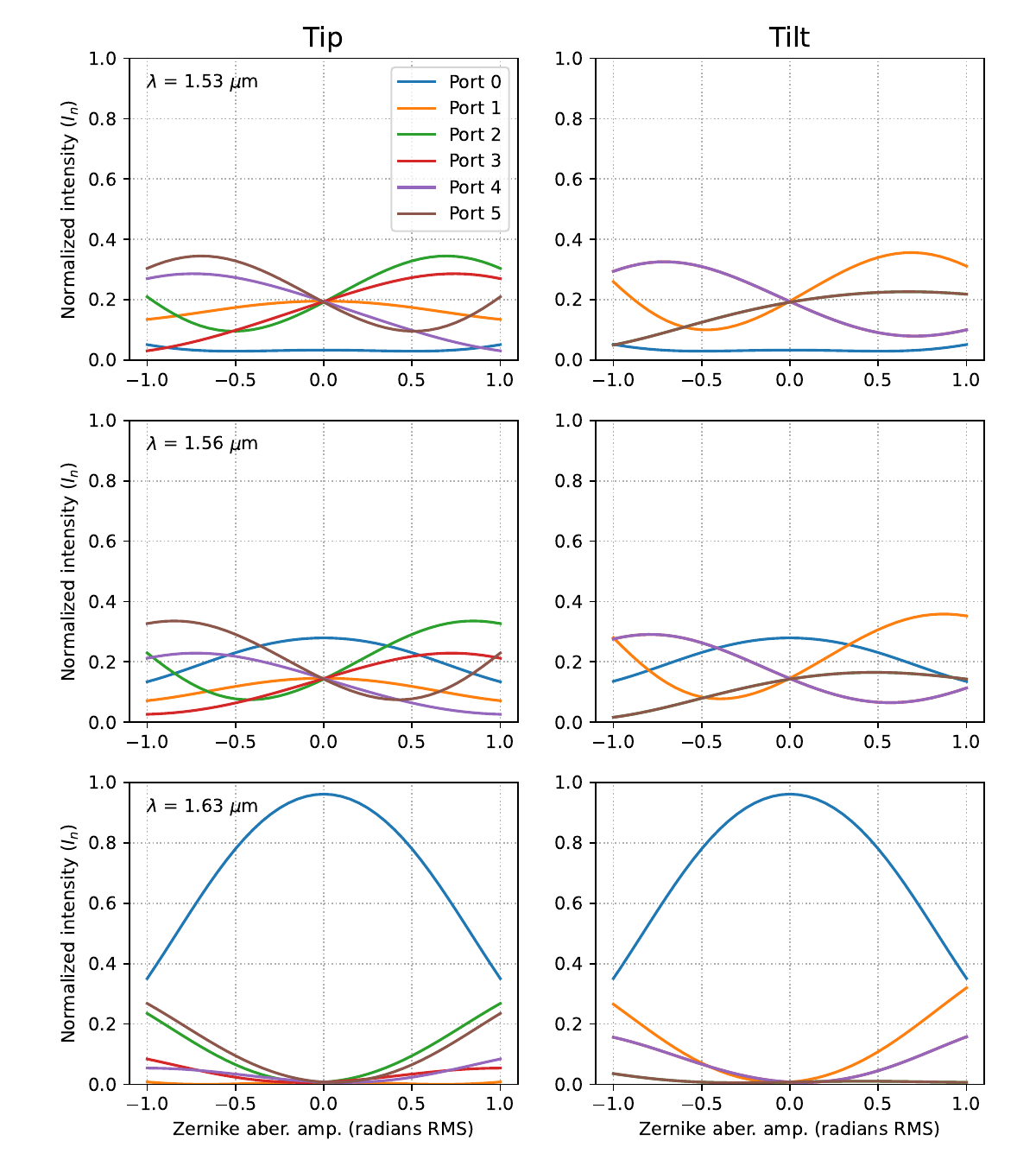}
\caption{Same as bottom panels of Figure \ref{fig:linearity} but with three different wavelengths. The intensity responses change slowly with wavelength. The oscillation of normalized intensity ($\mathbf{I}_{\rm n0}$) in the central port (port 0) and the rest and the oscillation of linear intensity responses ($B'_n$) are noticeable.}
\label{fig:chromaticity_6port}
\end{figure}

For the 6-port PL, the chromatic behaviors are more complicated because there is a greater number of modes (\LP{01}, \LP{11}, \LP{21}, \LP{02}) with differing propagation constants. 
Figure \ref{fig:chromaticity_6port} shows examples of normalized output intensities as a function of tip-tilt Zernike mode amplitudes for three different wavelengths. 
The zeropoints ($\mathbf{I}_{\rm n0}$) and the slopes ($B'_n$) vary as a function of wavelength. 
Tip-tilt sensitivity is determined by combined effects of relative phase between \LP{01} and \LP{11} modes and between \LP{02} and \LP{11} modes.
The relative phase between \LP{01} and \LP{02} modes as a function of wavelength causes oscillating behavior of normalized intensity in port 0 and the rest and sensitivity to defocus \cite{lin22}: the normalized intensity in port 0 being 
nearly zero at $\lambda = 1.53$\,\textmu m and nearly unity at $\lambda = 1.63$\,\textmu m. 
Consequently, the spectroastrometric S/N (Section \ref{ssec:photnoise}) and the jitter sensitivity (Section \ref{ssec:varyingWFE}) vary with wavelength. 
In our particular lantern case (taper length, geometry and refractive indices), the response to tip-tilt at  $\lambda=1.63$\,\textmu m, is highly nonlinear, resulting in larger jitter sensitivity and smaller spectroastrometric S/N. Hence, for accurate spectroastrometric measurements at the wavelengths of interest, the lantern properties should be carefully designed for optimum performance.

\section{Mock observation of accreting protoplanets}\label{sec:mockobs}

One of the possible applications of PL spectroastrometry is searching for companions with emission lines, such as accreting protoplanets around young stars \cite{whe15}.
Accreting planets are known to emit strong hydrogen emission lines because of heating in the accretion shock region \cite{aoy21,tak21}. The line luminosities are proportional to mass accretion rates \cite{aoy21}. Recently, H$\alpha$ direct imaging observations revealed accreting planets PDS 70b and PDS 70c around a young star PDS 70, at 113 pc, with angular separations around 200 mas \cite{wag18, haf19}. Spectroastrometry opens up the possibility of detecting such objects at much smaller separations. Moreover, the capability of two-dimensional spectroastrometry with PLs enables more efficient detection. 

We simulate simple mock observations of accreting planets around a PDS 70 analog as follows. The telescope and instrument throughputs are simulated using {\tt PSISIM} \cite{van23} \footnote{\url{https://github.com/planetarysystemsimager/psisim}}, assuming observations with the next generation high-resolution instrument at the W. M. Keck Observatory, High-resolution Infrared Spectrograph for Exoplanet Characterization (HISPEC)\cite{maw19,maw22}. 
We use {\tt PHOENIX} stellar atmosphere models \cite{hus13} to model the spectra of PDS 70, using an effective temperature $T_{\rm eff}=4000{\rm K}$, surface gravity $\log{g}=4.5$, and solar metallicity.
For the accreting planet, we generate Gaussian-shaped emission lines for two accretion rates $\dot M = 10^{-8}\, {\rm M_{\odot}}\, {\rm yr}^{-1}$ (case A) and $\dot M = 10^{-8.5}\, {\rm M_{\odot}}\, {\rm yr}^{-1}$ (case B), using accretion luminosity -- line luminosity scaling relations by Aoyama et al. \cite{aoy21}
We particularly focus on the hydrogen Paschen $\beta$ line (1.282 \textmu m) in $J$ band for this simulation.
The linewidth is set to 50${\rm km\, s^{-1}}$, corresponding to free-fall velocity of a Jupiter-mass object. 
We assume that the absolute magnitude of the host star is 4.2 magnitude in $J$ band. If the star is rapidly accreting ($\dot M > 10^{-8}\, {\rm M_{\odot}}\, {\rm yr}^{-1}$), the Paschen $\beta$ emission line will dominate over the continuum \cite{rig12}, but we do not consider the stellar emission line in this paper.
Then the contrasts between the star and the planet in Paschen $\beta$ line are $2.0\times10^{-2}$ and $5.2\times10^{-3}$, respectively. For case A, we assume a small separation (2 mas) at 40 pc which is in the linear intensity response regime. For case B we assume a larger separation (18 mas) at 140 pc distance, beyond the PL's linear intensity response regime but within $\lambda/D$ for $D = 10$~m, which we may constrain the separation and the contrast simultaneously (\S\ref{sssec:nonlinear}). 
Our simulated wavelength range is [1.278, 1.286] \textmu m. 
We ignore the chromatic behavior of the lantern (variation in transfer matrix as a function of wavelength) given that the wavelength range is small compared to the range of lantern chromatic behavior. We use the transfer matrix calculated for 1.55~\textmu m with lantern properties described in Section \ref{sssec:sim}. 
The spectra are binned to optimize the signal-to-noise ratio of the Paschen $\beta$ emission line.
For the effects of WFEs, we only consider wavelength-independent random tip-tilt jitter for both cases. 300 tip-tilt values are drawn from normal distribution with standard deviation of $\sigma_{\rm tt} = 0.2 \lambda/D$. 
Table \ref{tab:mockobs} lists all the simulation parameters.

\begin{figure}
    \centering
    \includegraphics[width=1\linewidth]{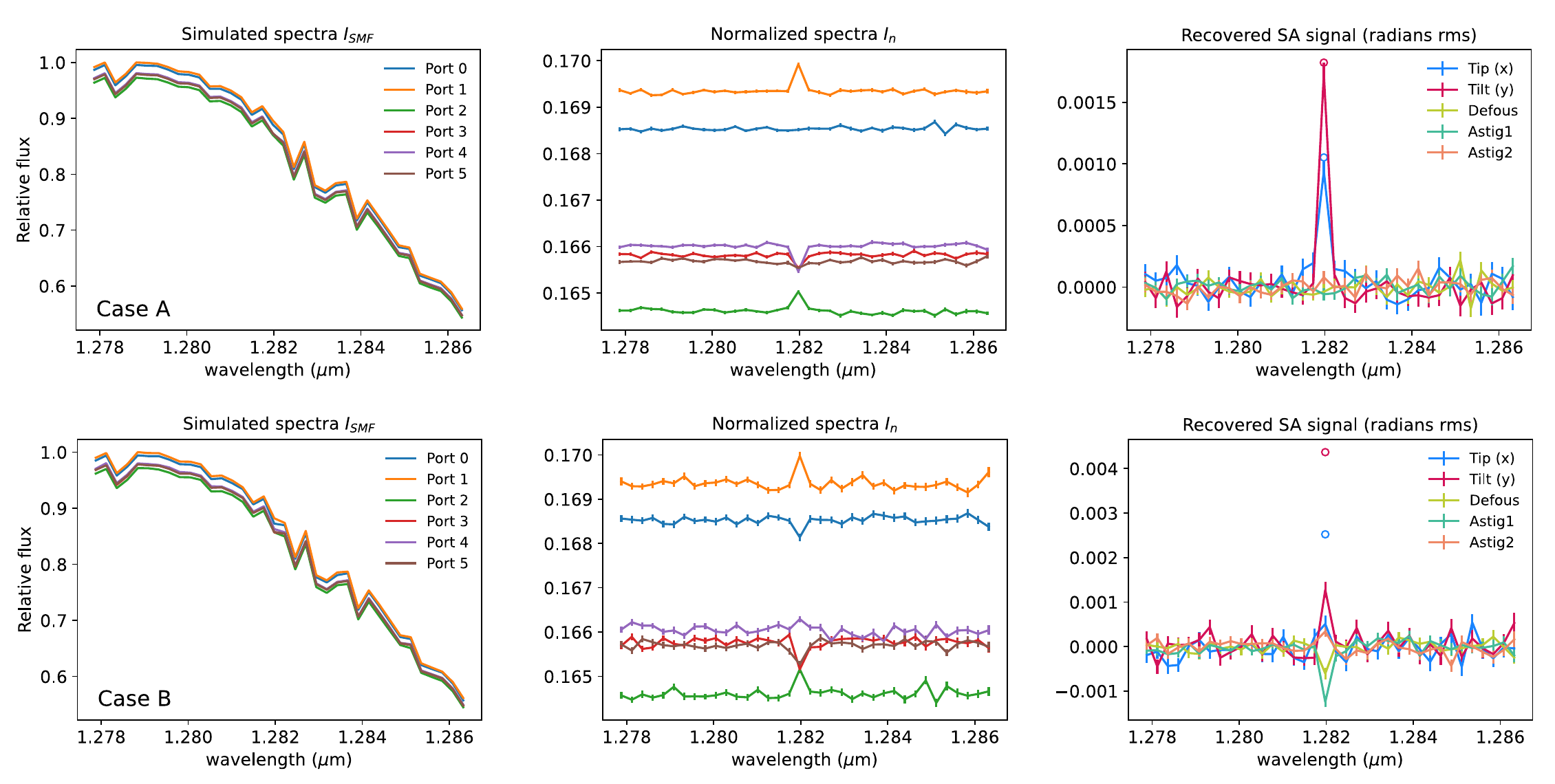}
    \caption{Example mock observations of an accreting protoplanet around a PDS 70 analog on Paschen $\beta$ line. The simulation parameters are described in Table \ref{tab:mockobs}. Top and bottom panels display the mock observations for the case A (linear intensity response regime) and case B (larger separation, nonlinear regime), respectively. (Left) Simulated spectra in 6 output SMFs. The spectral features are dominated by the absorption lines of the star. (Middle) The spectra normalized by the sum of intensities in the 6 ports as a function of wavelength. (Right) Recovered spectroastrometric signals as a function of wavelength. The signals from the off-centered emission line of the planet at the wavelength of Paschen $\beta$ are detected. The true center of lights are indicated as circular symbols. For case A, the recovered spectroastrometric signals match the true centroid shifts. For case B, the recovered tip-tilts deviate from the true centroid shifts
    but in this nonlinear regime, the detected normalized intensity signals can be used to fit a binary model to recover binary separation, contrast, and PA simultaneously.}
    \label{fig:demo_SA}
\end{figure}

\begin{figure}
    \centering
    \includegraphics[width=1\linewidth]{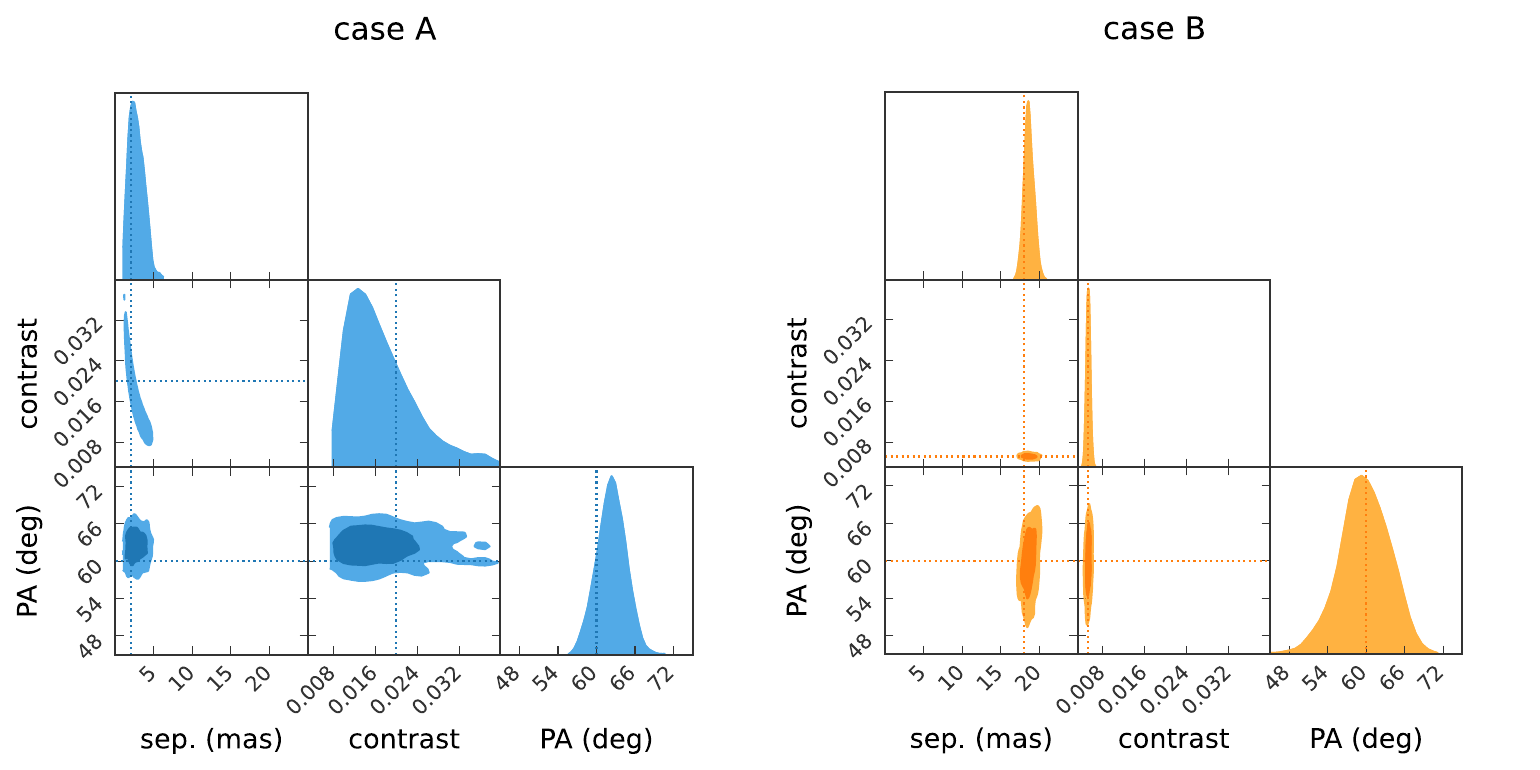}
    \caption{Inferred posterior distributions of the binary parameters for the case A (left) and case B (right). The true values of the parameters are indicated as dotted lines. The separation and contrast are degenerate for the small separation case (case A, in linear regime).}
    \label{fig:corner}
\end{figure}

The left panels in Figure \ref{fig:demo_SA} show simulated PL spectra for the case A (top) and case B (bottom), with photon noise and tip-tilt jitter. The output spectra are dominated by the stellar light with absorption features. 
The differences in the continuum levels between the ports reflect nonzero average tip-tilts. 
The middle panels show the normalized spectra $\mathbf{I}_{\rm n} (\lambda)$. In the normalized spectra, the signals from the off-center planet at $\lambda=1.282$\,\textmu m can be noticed. The right panels display the recovered spectroastrometric signals, Equation \ref{eq:centroid}, including defocus and astigmatism modes. 

For the case A where the separation is small (in linear regime), tip-tilt signals dominate. The non-detection in the astigmatism and defocus modes may imply that the planet separation is small. The recovered centroid shifts using the linear response matrix are close to true centroid shifts as indicated as circles in the right panel of Figure \ref{fig:demo_SA}. The recovered signals are systematically smaller than the true centroid shifts due to the second-order effects of the tip-tilt jitter (\S\ref{ssec:varyingWFE}).

For the case B where the separation is in nonlinear regime, there are noticeable astigmatism signals. In addition, the recovered spectroastrometric signals using the linear approximation fails as can be seen as the deviation of the recovered centroid shifts from the true centroid shifts. Although the S/N of tip-tilt modes are smaller in this case, detection in astigmatism modes can support the detection and provide constraints on separation, contrast, and PA simultaneously. 

As discussed in \S\ref{sssec:nonlinear}, we try model fitting the normalized spectra (middle panel of Figure \ref{fig:demo_SA}) to a binary model which can be constructed using empirical coupling maps.
In Table \ref{tab:mockobs} and Figure \ref{fig:corner} we display retrieved binary parameters and inferred posterior distributions using {\tt emcee} Markov Chain Monte Carlo (MCMC) ensamble sampler \cite{for13}. While the separation and contrast are highly degenerate for case A, due to nonlinearity, all the three binary parameters for case B are well-constrained despite lower S/N.

\begin{table}
\centering
\caption{Mock observation parameters and S/N of spectroastrometric signals} 
\label{tab:mockobs}
\begin{tblr}{
  hline{1,17} = {-}{0.08em},
  hline{2-3,11-12} = {-}{},
}
                                                & Case A               &                                     & Case B               &                                \\
Parameters                                      & Input                & Recovered                           & Input                & Recovered                      \\
Mass accretion rate ($M_{\odot}\, {\rm yr}^{-1}$) & $10^{-8}$            &                                     & $10^{-8.5}$          &                                \\
Distance (pc)                                   & 40                   &                                     & 140                  &                                \\
$J$~band stellar magnitude (mag)                & 7.2                  &                                     & 9.9                  &                                \\
Exposure time (hr)                              & 1                    &                                     & 3                    &                                \\
$\sigma_{tt}$                                   & 0.2                  &                                     & 0.2                  &                                \\
Separation (mas)                                & 2.0                  & $2.7^{+1.1}_{-0.9}$                 & 18                   & $18.6 \pm 0.6$                 \\
Contrast at Paschen~$\beta$                     & $2.0 \times 10^{-2}$ & $(1.6^{+0.8}_{-0.5})\times 10^{-2}$ & $5.2 \times 10^{-3}$ & $(5.2 \pm 0.4) \times 10^{-3}$ \\
Position angle (deg)                            & 60                   & $62 \pm 2$                          & 60                   & $60 \pm 4$                     \\
S/N                                             &                      &                                     &                      &                                \\
tip (x)                                         & 11                   &                                     & 3.4                  &                                \\
tilt (y)                                        & 19                   &                                     & 7.0                  &                                \\
defocus                                         & 0.8                  &                                     & 3.7                  &                                \\
astig1                                          & 0.9                  &                                     & 9.9                  &                                \\
astig2                                          & 0.6                   &                                     & 3.7                  &                                
\end{tblr}
\end{table}

\section{Discussion}\label{sec:discussion}

\subsection{Benefits of PL spectroastrometry}

Dispersed PL outputs enable two-dimensional spectroastrometry \cite{dav10, got12, mur13}, which 
enables a more efficient observation compared to long-slit spectroastrometry.
One of the advantages of PL spectroastrometry is that two-dimensional spectroastrometry can be easily enabled without the need to resample the focal plane, such as using image slicers. Once the focal plane field is coupled to the FMF entrance of a PL with a scale of about a few $\lambda/D$, small wavelength-dependent centroid shifts result in variations in the SMF output spectra, which can then be used to infer the centroid shifts. The centroid shifts can be efficiently recovered using a few-moded PL such as the 6-port PL described in this study. The few SMF outputs can be dispersed at high spectral resolutions \cite{lin21} with efficient use of the detector area. This can enable resolving two-dimensional kinematic structures that require high spectral resolution, as in the BLR and rotating star examples in \S\ref{sssec:scenes}.

Moreover, the capability of wavefront sensing with PLs \cite{nor20, lin22, lin23, lin23b} opens an extra potential to reduce the systematic effects of WFEs (\S\ref{ssec:varyingWFE}) by real-time wavefront correction and achieve a higher throughput. A PL filters out high-order aberrations and couples low-order aberrations that result in overall variations in SMF output intensities. The residual low-order aberrations may be corrected in real-time by using output intensities for active wavefront control.  
It may also be possible to characterize the low-order aberration effects in post-processing procedure, using chromaticity of the PL principal modes: the behaviors of the responses to the WFEs change slowly as a function of wavelength.

\subsection{Considerations on PL design}

To efficiently measure spectroastrometric signals, it is important that the PL has large linear intensity responses, $B_n$, which is directly related to the signal-to-noise ratio of the spectroastrometric signals (\S\ref{ssec:photnoise}). 
In addition, the susceptibility to systematic effects of time-varying WFE relies on the relative significance of the linear and cubic intensity responses (\S\ref{ssec:varyingWFE}).
Note that the intensity responses vary slowly as a function of wavelength, as discussed in \S\ref{sec:chromaticity}. It is essential for spectroastrometry to achieve a good linear response at the wavelength range of interest.
The intensity responses rely on the transfer matrix $A$ corresponding to the PLPMs, which can be designed and optimized for efficient spectroastrometric observations. By adjusting the geometry of the lantern such as changing the taper length and core arrangements, the intensity responses can be designed \cite{lin23}. 

Although we have limited our simulation to a standard 6-port PL in this paper, one may consider using a different mode-count PL or a different type of PL such as a mode-selective PL \cite{leo14}. If limiting consideration to the small separation regime (linear intensity response regime), a 3-port PL will be sufficient to recover tip-tilts. However, recovering spectroastrometric signals for a more extended object (angular extent in nonlinear intensity response regime) is more challenging due to higher degeneracy. 
A higher mode-count PL would exhibit a more complicated chromatic behavior, which may be more challenging to optimize in design. However, if the interest lies in characterizing a more extended object (\S\ref{sssec:nonlinear}), using a higher mode-count PL can be beneficial. Using a mode-selective PL, a deeper contrast may be achieved but with a position angle degeneracy of 180 degrees due to symmetry \cite{xin22}.

\section{Conclusion and future work}

In this work, we explored the capability of PLs for two-dimensional spectroastrometry. PLs can enable measuring two-dimensional spectroastrometric signals simultaneously without resampling of the focal plane, using a few single-moded spectral traces. We defined spectroastrometric signals for PLs and simulated them for a few simple scenes in \S\ref{sec:concept}. In the regime where the input scene has a small angular size such that PL's intensity responses are linear, the centroid shifts can be simply recovered using a linear response matrix. If the input scene is more extended, the PL relative intensities can be used to fit models to learn about more information other than centroid shifts. We investigated the effects of photon noise and WFEs on spectroastrometric signal recovery in \S\ref{sec:errors}. 
The spectroastrometric sensitivity for the photon noise-limited case is closely related to how linear the intensity responses are to tip-tilts and is theoretically comparable to conventional PSF centroid fitting for our simulated 6-port PL. The effects of static residual WFEs (wavelength-independent or slowly-varying as a function of wavelength) can be calibrated using continuum levels. The time-varying WFEs averaged over an exposure can affect the intensity responses, which may also be calibrated. We explored the chromaticity in PLPMs that result from phase differences between the waveguide modes that constitute PLPMs, in \S\ref{sec:chromaticity}. The linear intensity responses are expected to vary slowly as a function of wavelength, making some wavelengths more suitable for spectroastrometry than others. The design of the PL such as taper length can be optimized to achieve a good linear intensity response at the wavelength of interest.
We also provided mock observations of accreting protoplanets emitting hydrogen emission lines around a PDS 70 analog in \S\ref{sec:mockobs}.

Future work includes verifying the behavior of lanterns in a lab and on-sky, specifically the capability of measuring centroid shifts as a function of wavelength. 
The PL transfer matrices need to be experimentally determined and their stability under environmental changes such as temperature should be characterized (see Ref. \cite{lin23b} for stability of the response matrix in the context of PLWFS). The effects of two polarization states degenerate in SMF fundamental modes should be taken into account in the instrument design\cite{halverson15}. The PL transfer matrices will likely also depend on polarization at some level \cite{xin24}. Moreover, accurate port-by-port wavelength solutions are crucial for spectroastrometric measurements as well as reconstruction of the summed spectra, since wavelength solution errors can introduce artificial spectroastrometric signals and systematic broadening of the spectra. 
Practical observing strategies and calibration techniques to deal with alignment sensitivity and systematic effects regarding time-varying and static WFEs are also left for future investigation.

\subsection* {Code and Data Availability} 

The data and code used in preparation of this work are available upon request to the corresponding author.

\acknowledgments 
This work is supported by the National Science Foundation under Grant No. 2109231, 2109232, 2308360, and 2308361. An earlier version of this paper has been previously submitted as a SPIE conference proceeding \cite{kim22}. This research made use of Numpy\cite{harris2020array},  SciPy\cite{2020SciPy-NMeth}, HCIPy\cite{por18},  lightbeam\cite{lin21lightbeam},  Matplotlib\cite{Hunter:2007}, and pyGTC\cite{Bocquet2016}.


\bibliography{report} 
\bibliographystyle{spiebib} 


\end{document}